\documentclass[a4paper]{article}

\usepackage{INTERSPEECH2022}
\usepackage{subfig}
\usepackage{url}
\title{End-to-End Voice Conversion with Information Perturbation}
\name{Qicong Xie\thanks{*Work performed while interning at Tencent AI Lab.}$^{1,*}$, Shan Yang$^2$,  Yi Lei$^1$, Lei Xie$^1$, Dan Su$^2$}
\address{
  $^1$Audio, Speech and Language Processing Group (ASLP@NPU), School of Computer Science, \\ Northwestern Polytechnical University, Xi'an, China\\
  $^2$ Tencent AI Lab, China}
\email{xieqicong@mail.nwpu.edu.cn, leiyi@nwpu-aslp.org, lxie@nwpu.edu.cn, \{shaanyang, dansu\}@tencent.com}
\begin{document}

\maketitle
\begin{abstract}
The ideal goal of voice conversion is to convert the source speaker's speech to sound naturally like the target speaker while maintaining the linguistic content and the prosody of the source speech. However, current approaches are insufficient to achieve comprehensive source prosody transfer and target speaker timbre preservation in the converted speech, and the quality of the converted speech is also unsatisfied due to the mismatch between the acoustic model and the vocoder. In this paper, we leverage the recent advances in information perturbation and propose a fully end-to-end approach to conduct high-quality voice conversion. We first adopt information perturbation to remove speaker-related information in the source speech to disentangle speaker timbre and linguistic content and thus the linguistic information is subsequently modeled by a content encoder.  To better transfer the prosody of the source speech to the target, we particularly introduce a speaker-related pitch encoder which can maintain the general pitch pattern of the source speaker while flexibly modifying the pitch intensity of the generated speech. 
Finally, one-shot voice conversion is set up through continuous speaker space modeling. Experimental results indicate that the proposed end-to-end approach significantly outperforms the state-of-the-art models in terms of intelligibility, naturalness, and speaker similarity.

\end{abstract}

\vspace{-0.1cm}
\section{Introduction}
\vspace{-0.1cm}

The ideal purpose of a voice conversion system is to convert the source speaker's speech to make it sound naturally like the target speaker while maintaining the linguistic content as well as the speaking style of the source speech. With the development of deep learning, voice conversion has recently made significant progress~\cite{kaneko2017parallel,kaneko2018cyclegan,kameoka2018stargan,kameoka2018acvae}. However, the naturalness and speaker similarity of the converted speech is still not satisfactory. For low-resource scenarios, such as one-shot conversion, besides naturalness and speaker similarity, the conversation stability is another problem. 

Shaped as a one-dimensional time-series signal, human speech is rich in its content, including linguistic, para-linguistic and non-linguistic aspects~\cite{yamashita2013review}. The linguistic content mostly conveys language information to deliver, while prosody, speaking style or emotion are main para-linguistic aspects, and speaker timbre which mostly unveils speaker's identity is the most important non-linguistic aspect of speech. The key of voice conversion lies in the \textit{disentanglement} of the content information mentioned above. With successful disentanglement, we can easily transfer the linguistic as well as the para-linguistic (prosody/style/emotion) aspects of source speech to the voice of target speaker while keeping target speaker's timbre.

%linguistic content, speaker timbre, and prosodic information. 
%For voice conversion, it only needs to convert the timbre attribute of the source 
%speech to the target speaker, while maintaining the linguistic contents and the speaker style. So the main challenge is how to disentangle and model the above attributes. 

%However, there are still some problems with stability, naturalness, and speech quality due to unsatisfied content disentanglement and prosodic modeling, as well as the mismatch between the acoustic model and vocoder. The improvement of these aspects is of great significance for the widespread use of voice conversion technology. 

As for the disentanglement of the linguistic content, a common solution is to utilize a pre-trained speech recognition model to extract only text-related phonetic posteriorgrams (PPGs) of source speech~\cite{sun2016phonetic,wang2021enriching,wang2021one}. The PPGs can be considered as a speaker-independent intermediate representation to represent the linguistic content as the aim of a speech recognizer is to decode linguistic discriminant information from speech without the consideration of who is speaking.
Thus with speaker's timbre as additional information in a voice conversion model, the PPGs can be used as input to synthesize speech with target speaker's timbre~\cite{sun2016phonetic}. In addition to the linguistic content and speaker timbre information, the speaking style of source speech is also critical to the naturalness of the converted speech. However, with the target to remove linguistic-irrelevant information in the PPGs, it is hard to deliver the speaking style of source speech in the PPG-based voice conversion models, which makes it necessary to utilize an extra prosody model for providing the speaking style in the converted speech~\cite{wang2021enriching,wang2021one}.

 %always lack the pitch variations, it is hard to present the speaking style of the source speech in the PPG-based voice conversion models, which makes it necessary to utilize extra prosodic features for providing the speaking style~\cite{wang2021enriching,wang2021one}.

% Hence the target voice can be obtained with the help of additional timbre embedding~\cite{sun2016phonetic}. But the PPGs don't contain the pitch variation of the source speech, and there is no guarantee for maintaining the pitch trend of the source speech, which is typically predicted through the acoustic model. So we need to inject explicit prosodic features to maintain the source style~\cite{wang2021enriching,wang2021one}.

Besides, since the PPGs are extracted from an independent ASR model, the inevitable recognition errors will lead to mispronunciation in the converted speech. To solve this problem, there have been recent studies aiming at conducting voice conversion without the pre-trained ASR model. To directly extract linguistic information from the source speech during training, the existing models commonly try to disentangle the speaker timbre and linguistic content through different types of encoders. In AutoVC~\cite{qian2019autovc}, information bottleneck is learned to isolate speaker information from content information, while it still needs another pre-trained speaker encoder to further inject speaker identity. With the help of adaptive instance normalization~\cite{huang2017arbitrary},  speaker timbre and linguistic content can be separately modeled in a unified conversion model~\cite{chou2019one}. To further simplify the above solution, AGAIN-VC ~\cite{chen2021again} models the two aspects in a single encoder through activation guidance and adaptive instance normalization. But these methods usually treat the mel-spectrogram as the model target, where a vocoder is still required to reconstruct waveform. So there exists a potential mismatch problem, where the vocoding system models the distribution of mel-spectrogram from real speech while the predicted mel-spectrogram from the conversion model may have a different drifted distribution, affecting the quality of the synthesized speech. To mitigate this problem, a fully end-to-end voice conversion model named NVC-Net~\cite{nguyen2021nvc} is proposed to obtain high-quality converted speech, where the content preservation loss is utilized for disentangling the linguistic content. 

\begin{figure*}[htbp]
    \centering
    \includegraphics[scale=0.67]{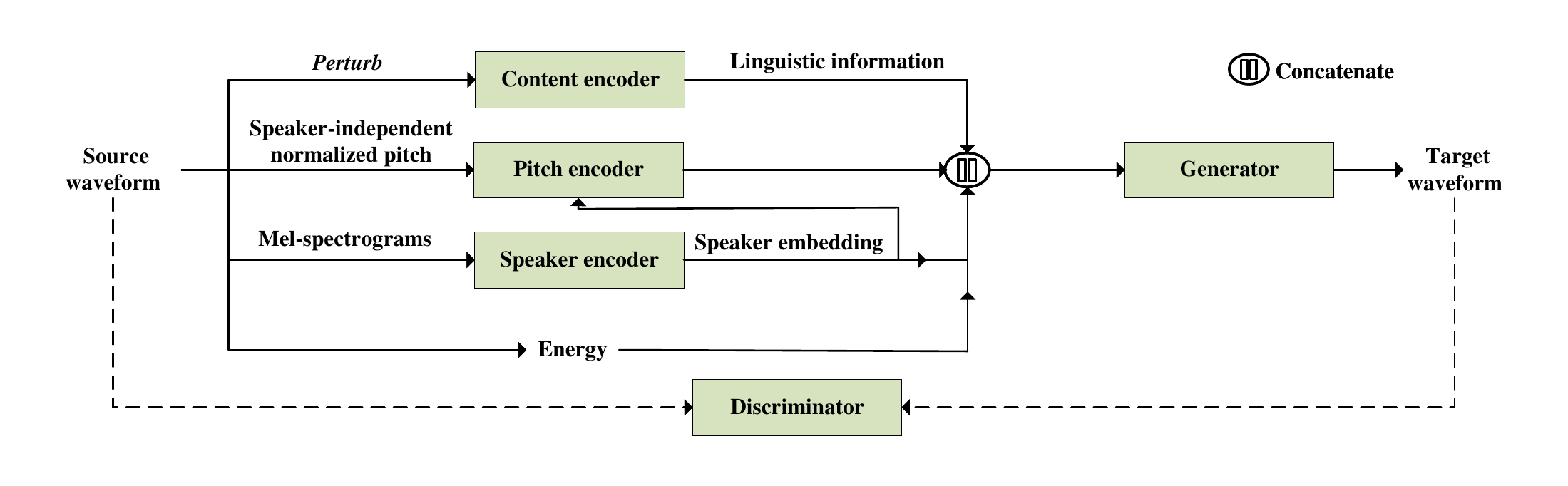}\vspace{-0.3cm}
    \caption{The architecture of the proposed model}\vspace{-0.3cm}
    \label{fig:model}
\end{figure*} 

\begin{figure}[!htb]
    % \label{fig:pitch}
  \begin{minipage}[t]{0.33\linewidth}
    \centering    
% \subfloat[content encoder]{\includegraphics[trim=20 0 20 0,scale=0.7]{IEEEtran/fig/content.pdf}}
\subfloat[content encoder]{\includegraphics[trim=20 0 15 0,clip,scale=0.7]{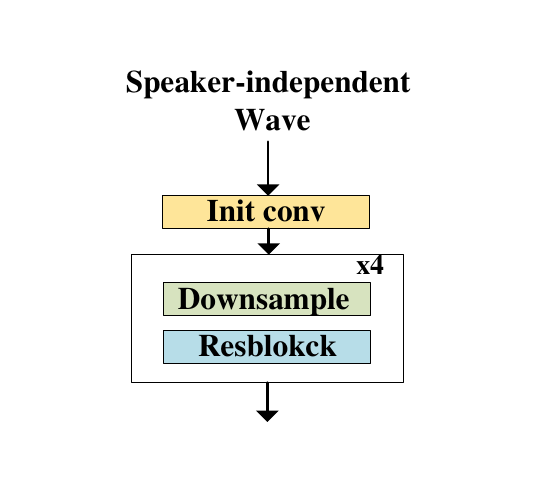}}
    % \label{fig:side:a}
  \end{minipage}%
  \begin{minipage}[t]{0.33\linewidth}
    \centering
\subfloat[speaker encoder]{\includegraphics[trim=15 0 35 0,clip,scale=0.7]{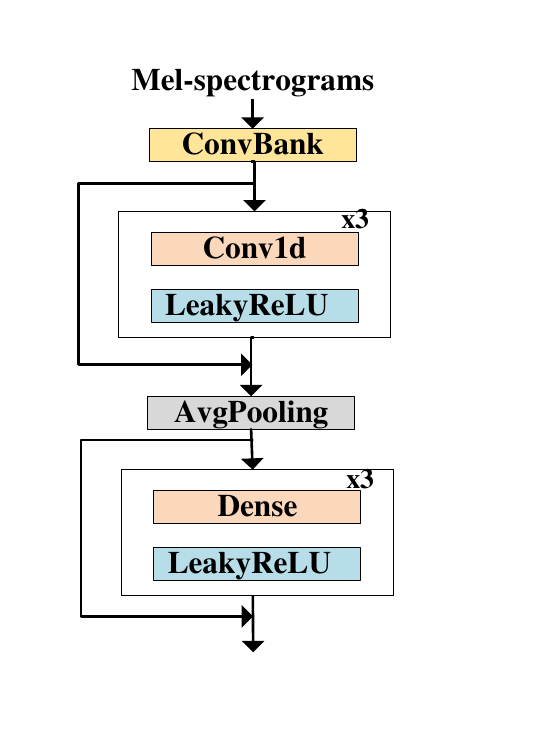}}
    % \caption{cap b}
    % \label{fig:side:b}
  \end{minipage}
    \begin{minipage}[t]{0.33\linewidth}
    \centering
\subfloat[pitch encoder]{\includegraphics[trim=35 0 20 0,clip,scale=0.7]{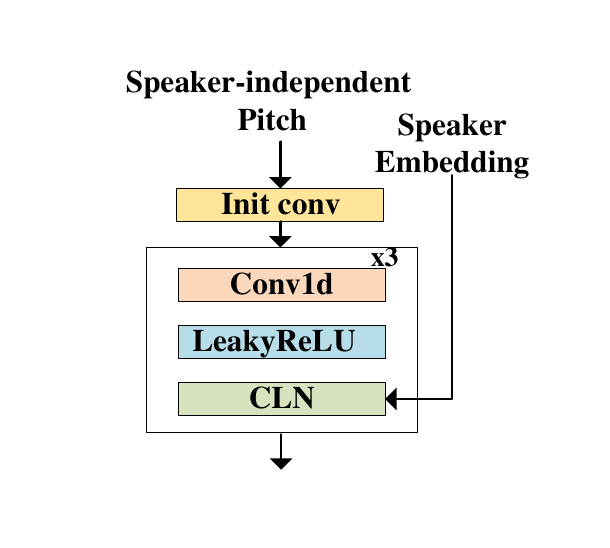}}
    % \caption{cap b}
    % \label{fig:side:b}
  \end{minipage}
  \vspace{-0.2cm} 
     \caption{The architecture of (a) content encoder, (b)speaker encoder, and (c) pitch encoder. ``CLN'' is abbreviation of ``conditional layer normalization''.}
    \label{fig:modelduraion}\vspace{-0.45cm} 
\end{figure}

% Both the above two-stage or end-to-end models conduct disentanglement on the source speech with either supervised or unsupervised manner. 
Although the above models can conduct disentanglement on the source speech and realize voice conversion, it still faces
the \emph{leakage of timbre} problem, i.e., the source speaker's timbre information is inevitably transferred into the synthesized speech, making it sound somehow like uttered by the source speaker or something between the two speakers. The main reason is attributed to the fact that the learned linguistic content embedding still contains some of the source speaker's timbre.

With the aim to fully remove speaker timbre in the source speech, \textit{information perturbation} has been recently adopted in the NANSY system~\cite{choi2021neural}. The basic idea of information perturbation is to simply perturb all the information we do not need in the source speech through signal processing, thereby the network will not extract the undesirable attributes from the inputs during training. 
In details, NANSY uses a pre-trained wav2vec~\cite{baevski2020wav2vec} model to extract linguistic representation and perturbs unwanted pitch the formant of the source speech. The additional pitch and energy constraints are further controlled in the decoder to product mel-spectrogram. 
Although NANSY achieves satisfied speaker similarity in the converted speech, it is still complex in system building and requires a vocoder to reconstruct the waveform.

Inspired by the advantage of information perturbation on speech attribute disentanglement, an information perturbation based fully end-to-end voice conversion model is proposed in this paper. Different from NANSY, in which the linguistic feature is extracted by a pre-trained wav2vec model, in the proposed method, we directly use a learnable network to extract linguistic information from the perturbed speech. Besides, as pitch is essential to speech prosody, we specifically introduce a speaker-related pitch modeling module to model the intonation aspect of prosody in the source speech. This module uses the speaker-independent normalized pitch as input and effectively transfers the pitch trend of the source speech to the target. Besides, the pitch module can flexibly modify the pitch (increase or decrease) of the generated speech while keeping the general pitch pattern transferred from the source speech. 
In order to further conduct one-shot voice conversion, we expand our model with a jointly learned speaker encoder, which can extract and provide target timbre for the conversion process.
% Without the mismatch problem in the two-stage paradigm, the proposed fully end-to-end structure is beneficial for high-quality converted speech.
Without the mismatch problem between the acoustic model and the vocoder, the proposed fully end-to-end structure is beneficial for high-quality converted speech.
Extensive experiments demonstrate that the proposed model outperforms the state-of-the-art baselines in terms of speech intelligibility, naturalness, and quality, and we can easily control the pitch expressions of the converted speech.

\vspace{-0.2cm}
\section{Proposed Approach}
\vspace{-0.1cm}

\subsection{System overview}

Figure~\ref{fig:model} shows the general structure of the proposed model, which contains a content encoder, a pitch encoder, a speaker encoder, a waveform generator, and a discriminator. The content encoder tends to extract only linguistic-relevant features from the perturbed waveform, which is speaker-independent. To further inject speaker identities for waveform reconstruction, we adopt a speaker encoder to extract the timbre of source speech. Since the prosodic features like pitch are critical for the naturalness of the synthesized speech in voice conversion, we further propose a speaker-dependent pitch modeling module to maintain the pitch variation trend of the source speaker and adjust the pitch range of the target speaker. With the control of the learned linguistic content representations, timbre information, pitch embedding, and additional energy, the waveform generator aims at reconstructing target speech. Finally, we apply the adversarial training in our model to improve the quality of synthesized speech.

\vspace{-0.1cm}
\subsection{Content modeling through information perturbation}
The goal of the content encoder is to extract the only linguistic-related information from the source speech and remove the speaker-related information. Motivated by~\cite{choi2021neural},we employ three perturbation functions, including  formant shifting ($fs$)~\cite{jadoul2018introducing}, pitch randomization ($pr$)~\cite{jadoul2018introducing}, and random frequency shaping using a parametric equalizer ($peq$)~\cite{zavalishin2012art}, to remove the speaker-related aspects for unsupervised linguistic content extraction. The formant shifting function can modify the formant of the source speech, which is highly related to the timbre. Besides, the pitch randomization function is also adopted to avoid speaker-dependent pitch range. And the function $peq$ aims at conducting random frequency shaping, which consists of a series of low-shelving, peaking, and high-shelving filters:
\begin{equation}
% \vspace{-0.1cm}
H^{\mathrm{PEQ}}=H^{\mathrm{LS}} H^{\mathrm{HS}} \prod_{i=1}^{8} H_{i}^{\text {Peak }},
% \vspace{-0.1cm}
\end{equation}
where $H^{\mathrm{LS}}$ is a low-shelving filter, $H^{\mathrm{Peak}}$ stands for a peaking filter, and $H^{\mathrm{HS}}$ is a high-shelving filter, each of which  is a second-order IIR filter. The cutoff frequencies of $H^{\mathrm{LS}}$ and $H^{\mathrm{HS}}$ are fixed at 60 Hz and 10 kHz, respectively. The center frequency of the $H_{i}^{\text {Peak }}$ are evenly distributed among the shelving filters on a logarithmic scale. Based on the above perturbation function, we perturb the source speech $x$ through:

\begin{equation}
% \vspace{-0.1cm}
\hat x = fs(pr(peq(x))),
% \vspace{-0.1cm}
\end{equation}
where the output $\hat x $ doesn't maintain the speaker-related information of $x$. In this way, we argue that the latter content encoder will only focus on learning the speech linguistic contents.

The content encoder in the proposed model is a fully convoluted neural network, which encodes the perturbed audio into a downsampled content representation. As shown in Figure~\ref{fig:modelduraion} (a), the speaker-independent $\hat x$ firstly goes through a preprocessing convolution layer and then is fed into four residual blocks with down-sampling module. The architecture of the residual block is similar to that of  ~\cite{kong2020hifi}, where leaky ReLU activation and weight normalization are applied for all convolution layers.

\vspace{-0.1cm}
\subsection{Timbre and pitch modeling}

Since the content encoder only aims at extracting speech linguistic content, we also inject speaker information for further waveform reconstruction. As shown in Figure~\ref{fig:modelduraion} (b), we utilize a speaker encoder to extract speaker embedding from the mel-spectrogram. The final speaker representation is obtained through an average pooling layer and three fully connected layers with residual connections.

In voice conversion, we tend to maintain the pitch variation trend of the source speaker, which is performed as the target speaker. To this end, we proposed a speaker-related pitch module to consider both the pitch variation and the speaker identity, which aims at converting the speaker-independent source pitch trend into the pitch range of the target speaker. In details, we first calculate the speaker-level mean and variance of each speaker, and then normalize the source pitch to obtain a speaker-independent pitch. Then the normalized pitch and the speaker embedding are utilized to learn the final pitch embedding through the pitch encoder, in which way the pitch values of the reconstructed speech are defined by both the source pitch trend and the target speaker embedding.

As shown in Figure~\ref{fig:modelduraion} (c), the pitch encoder consists of a convolution layer followed by three conditional encoding blocks, each of which includes a convolutional layer with ReLU activation and a conditional layer normalized convolutional layer~\cite{chen2021adaspeech}. The conditional layer takes the speaker embeddings representation to model the speaker constraint.

\vspace{-0.1cm}
\subsection{Controllable waveform reconstruction}
Integrating the learned linguistic content representation, speaker information, and pitch embedding, we then set up a waveform generator with additional energy. In this way, we can easily control different attributes like timbre, linguistic content and pitch of the outputs to obtain controllable synthesized speech. As for the end-to-end voice conversion, we can easily extract the linguistic contents and the pitch trend of the source speech, and then obtain the target converted speech with a specific speaker embedding.

The architecture of the generator inherits from HiFiGAN~\cite{kong2020hifi}, which consists of four upsampling modules with residual blocks. To further improve the quality of synthesized speech, we also apply the adversarial training on the predicted waveform, where the multi-period discriminator (MPD),  multi-scale discriminator (MSD) and the multi-resolution spectrogram discriminator are adopted in the discriminative process~\cite{kong2020hifi,jang2020universal}. So the training objective of the proposed model is:
\begin{equation}
\mathcal{L_G}=\mathcal{L}_{a d v_g}+\mathcal{L}_{fm}+\mathcal{L}_{stft}
\end{equation}
\begin{equation}
\mathcal{L_D}=\mathcal{L}_{a d v_d}
\end{equation}
where $\mathcal{L}_{a d v_g}$ and $\mathcal{L}_{adv_D}$ are the adversarial loss of the generator and discriminator respectively, and $\mathcal{L}_{fm}$ is the feature matching loss as described in~\cite{yamamoto2020parallel}. Besides, the multi-resolution STFT loss $\mathcal{L}_{stft}$~\cite{yamamoto2020parallel} is also applied to improve the training stability and audio quality.

% \subsection{Training and generation}
\vspace{-0.2cm}

\section{Experiments}
\vspace{-0.1cm}

\subsection{Basic Setup}
All our experiments are performed on the VCTK~\cite{veaux2017supersededvctk} dataset, which contains speech data from 109 English speakers with different accents. For data preprocessing, all speech sentences are downsampled to 24 kHz, and the silence at the beginning and end of the audio was cut off. For evaluation, we randomly reserve three male and three female speakers as unseen speakers.

We set up two state-of-the-art one-shot voice conversion models, AGAIN-VC~\cite{chen2021again} and NVC-Net~\cite{nguyen2021nvc}, as baseline systems to evaluate the proposed model. The AGAIN-VC applies a single encoder for disentangling speaker and linguistic content information, while the NVC-Net utilizes the content preservation loss for disentanglement. During training, we randomly perturb the input speech at each step for the content encoder, while the pitch and energy are extracted from the raw waveform. We use Adam~\cite{kingma2014adam} optimizer to optimize our model with a learning rate of 2e-4, $\beta_1$ = 0.8 and $\beta_2$ = 0.99 using a batch size of 64 until convergence. Both the proposed model and the baselines are trained on 4 NVIDIA V100 GPU, where the baseline models are conducted with their official implementations.

For evaluation, we conducted Mean Opinion Score (MOS) tests to evaluate the synthesized results in terms of speaker similarity and speaker naturalness. For better analysis, we individually conduct evaluation in four scenarios, including seen2seen, seen2unseen, unseen2seen, and unseen2unseen, where seen2unseen means the source speaker is seen during training and the target speaker isn't. There are total 20 listeners participating in each MOS test,
and in each test scenario, we generate 10 utterances for each of the six target speakers. The audio samples can be found at our demo page\footnote{\url{https://qicongxie.github.io/end2endvc/}}.

\begin{table*}[!htbp]
 \caption{Comparison of our proposed method with AGAIN-VC and NVC-Net in terms of speaker similarity and speaker naturalness MOS with confidence intervals of 95$\%$ under 4 voice conversion scenarios. The higher value means better performance, and the bold indicates the best performance out of three models in each scenario.}
 \label{tab:transfer}
 \vspace{-0.2cm}
\setlength{\tabcolsep}{3mm}
 \centering
\begin{tabular}{l|lll|lll}
\toprule
\multicolumn{1}{c|}{} & \multicolumn{3}{c|}{Speaker similarity}                                                  & \multicolumn{3}{c}{Speaker naturalness}                                                \\ \cmidrule{2-7} 
\multicolumn{1}{c|}{}                         & \multicolumn{1}{c}{AGAIN-VC~\cite{chen2021again}} & \multicolumn{1}{c}{NVC-Net~\cite{nguyen2021nvc}} & \multicolumn{1}{c|}{Proposed} & \multicolumn{1}{c}{AGAIN-VC~\cite{chen2021again}} & \multicolumn{1}{c}{NVC-Net~\cite{nguyen2021nvc}} & \multicolumn{1}{c}{Proposed} \\ \midrule
seen2seen             & \multicolumn{1}{c}{3.12$\pm$0.047}  & \multicolumn{1}{c}{3.89$\pm$0.044} & \multicolumn{1}{c|}{\bf{3.97$\pm$0.038}} & \multicolumn{1}{c}{2.84$\pm$0.053}  & \multicolumn{1}{c}{3.69$\pm$0.047} & \multicolumn{1}{c}{\bf{4.08$\pm$0.033}} \\
unseen2seen            & \multicolumn{1}{c}{3.02$\pm$0.062}  & \multicolumn{1}{c}{3.55$\pm$0.073} & \multicolumn{1}{c|}{\bf{3.62$\pm$0.063}} & \multicolumn{1}{c}{2.51$\pm$0.054}  & \multicolumn{1}{c}{3.50$\pm$0.062} & \multicolumn{1}{c}{\bf{3.71$\pm$0.052}} \\
seen2unseen               & \multicolumn{1}{c}{2.98$\pm$0.062}  & \multicolumn{1}{c}{3.51$\pm$0.068} & \multicolumn{1}{c|}{\bf{3.57$\pm$0.061}} & \multicolumn{1}{c}{2.48$\pm$0.046}  & \multicolumn{1}{c}{3.59$\pm$0.051} & \multicolumn{1}{c}{\bf{3.84$\pm$0.044}} \\
unseen2unseen            & \multicolumn{1}{c}{2.56$\pm$0.056}  & \multicolumn{1}{c}{3.23$\pm$0.044} & \multicolumn{1}{c|}{\bf{3.34$\pm$0.050}} & \multicolumn{1}{c}{1.90$\pm$0.046}  & \multicolumn{1}{c}{3.02$\pm$0.036} & \multicolumn{1}{c}{\bf{3.30$\pm$0.042}} \\  \midrule
overall           & \multicolumn{1}{c}{2.92$\pm$0.029}  & \multicolumn{1}{c}{3.55$\pm$0.027} & \multicolumn{1}{c|}{\bf{3.63$\pm$0.026}} & \multicolumn{1}{c}{2.43$\pm$0.029}  & \multicolumn{1}{c}{3.45$\pm$0.026} & \multicolumn{1}{c}{\bf{3.73$\pm$0.022}} \\ \bottomrule
\end{tabular}
\vspace{-0.2cm}
\end{table*}

\vspace{-0.1cm}
\subsection{Speaker Similarity Evaluation}

We firstly evaluate the speaker similarity of converted speech from different systems in all scenarios, as shown in  Table~\ref{tab:transfer}. From the results, we can find the proposed method consistently outperforms the two baseline models in all scenarios, which indicates the effectiveness of the information perturbation based speech disentanglement. But in details, when the target speaker is seen during training, which refers to any2many VC scene, we find the speaker similarity is much higher for the seen source speakers for all end-to-end systems. We argue that in the end-to-end voice conversion, the input waveform still exists residual timbre information in above models, which affects the similarity of the synthesized speech. As for the one-shot voice conversion, also named any2any VC, the speaker similarity also decreases when the target speaker is unseen. We assume a pre-trained speaker encoder with larger amount of speakers may improve the performance for unseen targets.

\vspace{-0.1cm}
\subsection{Speech Naturalness Evaluation}
We also evaluate the naturalness of the converted speech for different models, as shown on the right side of Table~\ref{tab:transfer}. Similar to the results of speaker similarity, the proposed model outperforms the two baseline models in terms of the naturalness. Meanwhile, the MOS scores of our model and the NVC-Net are significantly higher than AGAIN-VC, which benefits from the fully end-to-end structure. Since the fully end-to-end model could directly predict the speech samples, which avoids the feature mismatch between training and inference caused by the vocoder and significantly improve the speech quality. In addition, the proposed model is better than the NVC-Net in naturalness, since the proposed model could transfer the prosody of the source speech to the target speaker, which makes the converted speech more natural.

\vspace{-0.1cm}
\subsection{Pitch Correlation and Control}
\begin{figure}[!htb]
 \vspace{-0.2cm}
  \begin{minipage}[t]{0.5\linewidth}
    \centering    
    
\subfloat[pitch trend ]{\includegraphics[trim=15 0 15 0,scale=0.27]{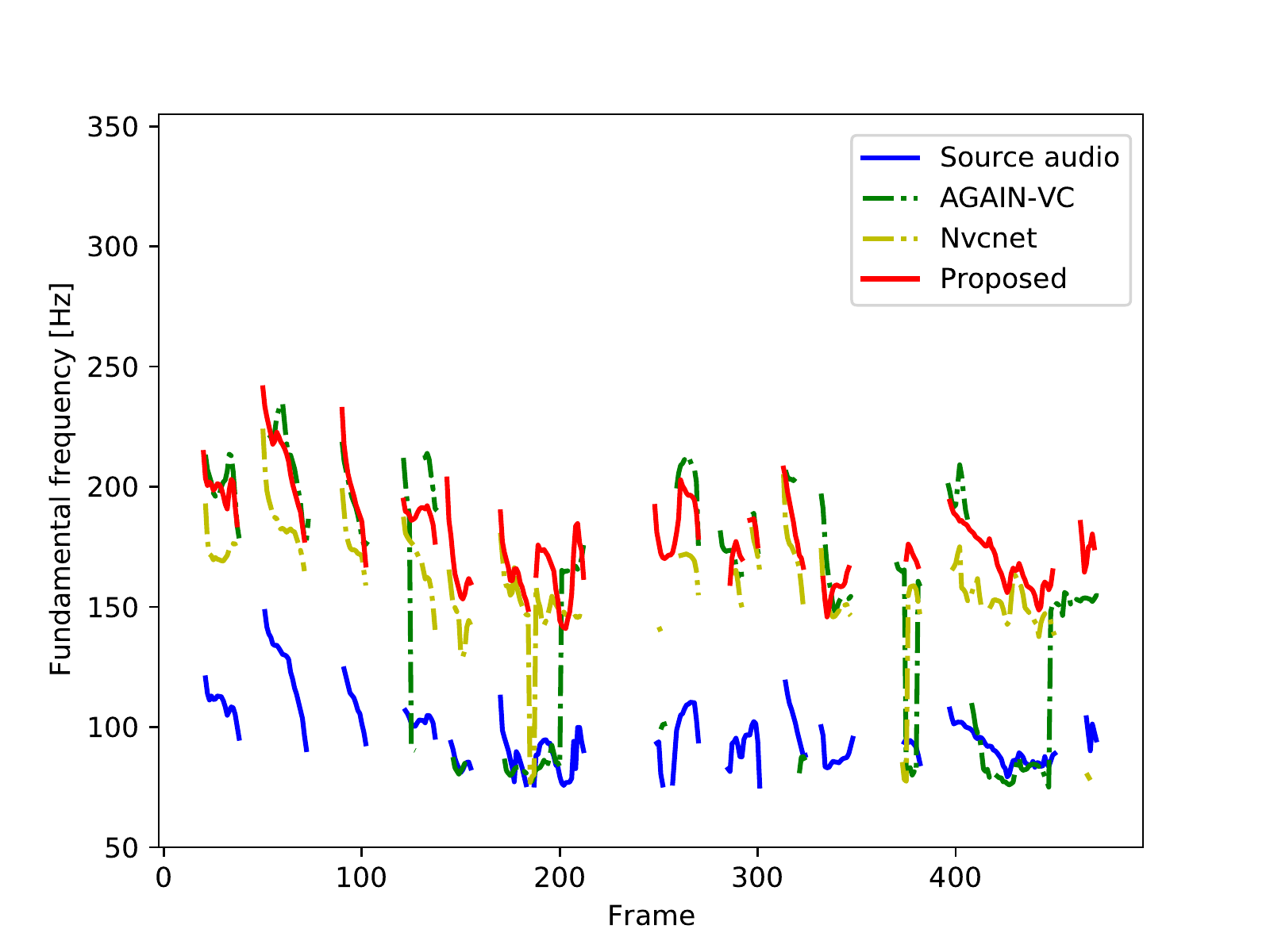}}
    \label{fig:side:a}
  \end{minipage}%
  \begin{minipage}[t]{0.5\linewidth}
    \centering
\subfloat[pitch control]{\includegraphics[trim=15 0 15 0,scale=0.27]{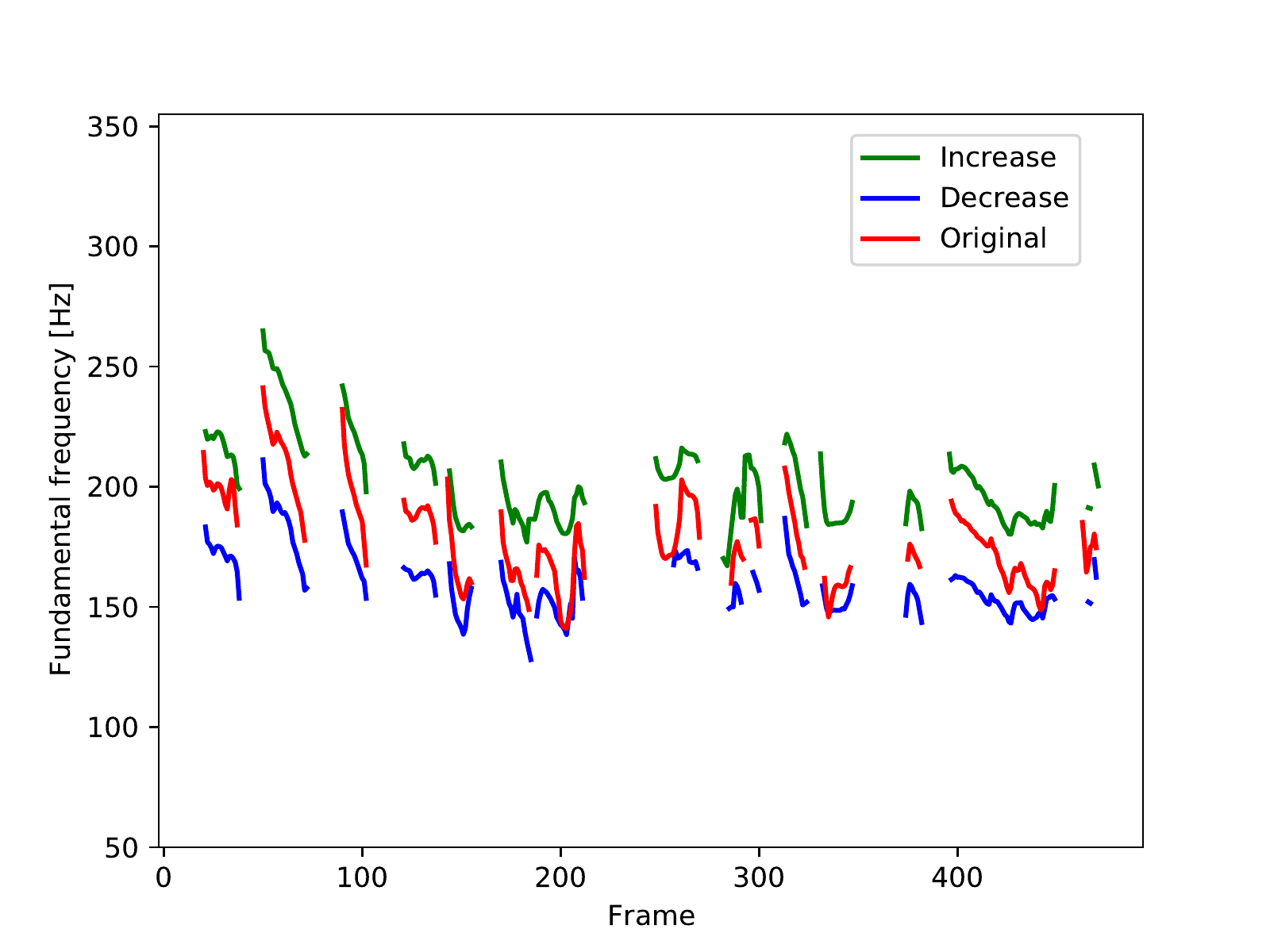}}
    % \caption{cap b}
    \label{fig:side:b}
  \end{minipage}
     \caption{(a) Pitch values of the source speech and the converted speech from three models. (b) Pitch of converted speech from proposed model with control factors.}\vspace{-0.3cm}
    \label{fig:pitchtrend} 
\end{figure}

To further demonstrate the prosody modeling ability of the proposed model, we convert a male source speaker's utterance into a female speaker and visualize the pitch of the speech from different models, as shown in figure~\ref{fig:pitchtrend} (a). It can be observed that the generated speech from three models all has higher pitch values lying in the target female speaker's pitch range. But compared to the baseline models, we can find that the pitch trend of the converted speech from the proposed model is the most similar to that of the source speech. It indicates that the proposed model could maintain the prosody of the source speech well.

To further confirm the disentanglement ability of the proposed model, we also modify the input pitch of the source speech to see the pitch changes in the converted speech. Figure~\ref{fig:pitchtrend} (b) shows the pitch values of the converted speech of our model by decreasing and increasing the input pitch. The results show that although the pitch values of the three sentences have changed, they still share the same pitch variation from the source speech.

\vspace{-0.1cm}
\subsection{Speech Intelligibility Evaluation}
To evaluate the intelligibility of the converted speech from the three models, we compute the Word Error Rate(WER) of the converted speech. To obtain WER, we utilize a pre-trained ASR model from  WeNet~\cite{yao2021wenet}, which is trained with the librispeech~\cite{panayotov2015librispeech} corpus. For evaluation, we randomly select 1000 sentences as source speech, which are converted into randomly selected six seen speakers and the reserved six unseen speakers. The WERs of the source speech and the converted speech are shown in Table~\ref{tab:compare_wer}, where we treat the WER of source speech, named GT is the topline of the ASR model. The results indicate that the proposed model achieves the lowest WER, which is significantly better than the two baseline models.  It illustrates that the proposed content encoder through information perturbation could effectively model the linguistic contents of the source speech.

\begin{table}[!htb]
\vspace{-0.1cm}
 \caption{Comparison of our proposed method with AGAIN-VC and NVC-Net in terms of WER, where the GT means the WER of the ground-truth source speech.}\vspace{-0.1cm}
 \label{tab:compare_wer}
  \centering
\vspace{-0.1cm}

\begin{tabular}{ccccc}
\hline

    & GT & AGAIN-VC & NVC-Net & Proposed       \\ \hline
WER & 7.24        & 34.82    & 28.76  & \textbf{12.89} \\ \hline
% cer & 5.04        & 24.86    & 20.41  & \textbf{8.07}  \\ \hline
\end{tabular}
\end{table}

\vspace{-0.2cm}
\section{Conclusions}
\vspace{-0.1cm}

This paper proposes a novel fully end-to-end high-quality voice conversion model with information perturbation. We utilize several perturbation functions to remove the speaker-related information in the source speech to disentangle the timbre and the linguistic contents, which avoids the leakage of timbre from the source speech. Besides, we propose a novel speaker-dependent pitch module to maintain the speaking style of the source speaker in the converted speech, making the generated speech more natural. Furthermore, we extend the proposed model to one-shot voice conversion scenario with a learnable continuous speaker space. The experimental results indicate that the proposed model outperforms the state-of-the-art voice conversion systems in terms of intelligibility, naturalness, and speaker similarity.

%The proposed model could We utilize information perturbation to remove the speaker-related information in the source speech so as to disentangle the timbre and content information without extra pre-trained models, which can improve timbre leakage and intelligibility problems. Besides, our model introduces a speaker-dependent pitch modeling module to convert the prosodic trend of the source speaker into the target domain, making the synthesized speech sound more natural. Furthermore, our proposed model can conduct one-shot voice conversion through a learnable speaker encoder. The fully end-to-end structure avoids potential feature mismatch problems caused by vocoder and improves the quality of synthesized speech. Experiment results indicate that our proposed system outperforms the state-of-the-art one-shot voice conversion systems in terms of quality, naturalness, and speaker similarity.

\vfill\pagebreak

\bibliographystyle{IEEEtran}

\bibliography{mybib}

\end{document}